# Intrinsic Electrochemical Limits Preceding Dendrite Penetration in Ceramic Electrolytes

*Rajeev Gopal[a] and Peng Bai[a,b,*]*

[a] Department of Energy, Environmental and Chemical Engineering, Washington University in St. Louis, 1 Brookings Dr, St. Louis, MO 63130, USA.

[b] Institute of Materials Science and Engineering, Washington University in St. Louis, 1 Brookings Dr, St. Louis, MO 63130, USA.

* Correspondence to: pbai@wustl.edu

**Abstract**

Solid-state electrolytes have the potential to stabilize lithium metal anodes, which hold the promise to nearly double the energy density of lithium-ion batteries. However, lithium metal dendrites penetrate the solid-state electrolyte during battery charging, undermining efficiency, limiting cycle life, and rendering high safety risks in practical applications. While multiple explanations have been proposed, understandings of the dynamics preceding and leading to the onset of metal penetration through solid electrolytes are still not conclusive. Here, by testing over 60 highly consistent $Li_{6.5}La_3Zr_{1.5}Ta_{0.5}O_{12}$ (LLZTO) samples with a unique *operando* technique, we reveal intrinsic electrochemical characteristics of current-dependent and thickness-dependent dendrite *initiation* dynamics, which are unifiable with the generalized Sand's time theory for binary liquid electrolytes. Our study explains and highlights the significance of long-range transport in ceramic electrolytes that are responsible for metal dendrite initiation.

# 1. Introduction

The increasing demand for high energy density batteries has sparked interest in all-solid-state batteries[1], in which a thin layer of lithium metal anode is coupled with a solid-state electrolyte (SSE) to not only increase the energy density by removing the bulky intercalation anode, but also enhance the safety by eliminating the use of flammable liquid electrolytes. Various Li-ion-conducting SSEs exhibiting ionic conductivity close to that of liquid electrolytes have been developed[2–5], among which the garnet-type cubic $Li_7La_3Zr_2O_{12}$ (LLZO) is one of the most promising candidates due to its excellent stability with lithium metal[6,7]. The hard and stiff oxide ceramic electrolyte is also expected to block any soft lithium metal dendrites, which plagued the development of Li metal batteries using liquid. However, dendritic lithium metal penetration still occurs in LLZO[9–11] when a charging current density higher than the empirical critical current density (CCD) is applied.

Recent advancements in understanding the dynamics of lithium dendrite formation in SSEs converge to three major causes that can occur concurrently: the current-focusing at the solid-solid interface due to void formation[12–15], the interplay between crack propagation with lithium metal filling[16–18], and the remote lithium metal nucleation due to electronic conductivity[19,20]. However, these interpretations cannot explain the thickness-dependent CCD, that thinner SSEs allow much higher charging current density, as demonstrated in thin-film LLZO electrolytes[21,22] and lithium phosphorus oxynitride (LiPON) electrolytes[20]. As we shall see in this study, the existence of current-dependent "incubation" time, i.e. from the start of galvanostatic to the onset of inward metal growth, cannot be attributed to interfacial or mechanical mechanisms. Instead, long-range transport and concentration polarization appear to play important roles.

In this work, we chose the Ta-doped cubic $Li_{6.5}La_3Zr_{1.5}Ta_{0.5}O_{12}$ (LLZTO) as a model system and examined over 60 highly consistent miniature Li|LLZTO|Li cells using *operando* video microscopy under one-way galvanostatic conditions. We reveal three electrochemical characteristics of lithium dendrite initiation in LLZTO, which match with those found in liquid electrolytes: (1) the critical current density to induce dendrite penetration is a system-specific (thickness-dependent) characteristic current density; (2) the current-dependent "incubation time" found in SSEs is consistent with the classic Sand's time scaling, and (3) the transition of the growth dynamics from outward growth to the inward growth always occurs at the characteristic incubation time. Our findings suggest the existence of significant concentration polarization due to transport limitation preceding the onset of lithium dendrite penetration through SSEs. The understanding of dendrite *initiation* dynamics in SSEs, i.e. before the nucleation of lithium metal within the SSEs, may therefore be unified with the classical understandings for dendrite initiation in liquid electrolytes.

## 2. Results

### 2.1 *Operando* Visualization at Varying Current Densities

To better understand the influence of current density on dendrite initiation and propagation in LLZTO, we performed *operando* experiments with tiny Li|LLZTO|Li symmetric cells constructed within glass capillaries for improved observation and ideal cell contact and alignment. As we shall see later, our miniature cells also mitigate the current-focusing effect, allowing a more uniform current distribution. For the LLZTO sample, our cold-press and sintering process[23] enabled a high relative density of > 95%, giving a translucent tan coloration. The surfaces were polished parallel and absent of any obvious flaws before lithium was applied. This setup allowed the application of pressure via the current collectors to mitigate potential interface delamination[24,25] and void formation[14,26], while simultaneously allowing the capture of any visual phenomena within the LLZTO to be correlated with the accompanying voltage response. The *operando* videos were post-processed to better display any changes (Figure 1) and synchronized with the voltage responses.

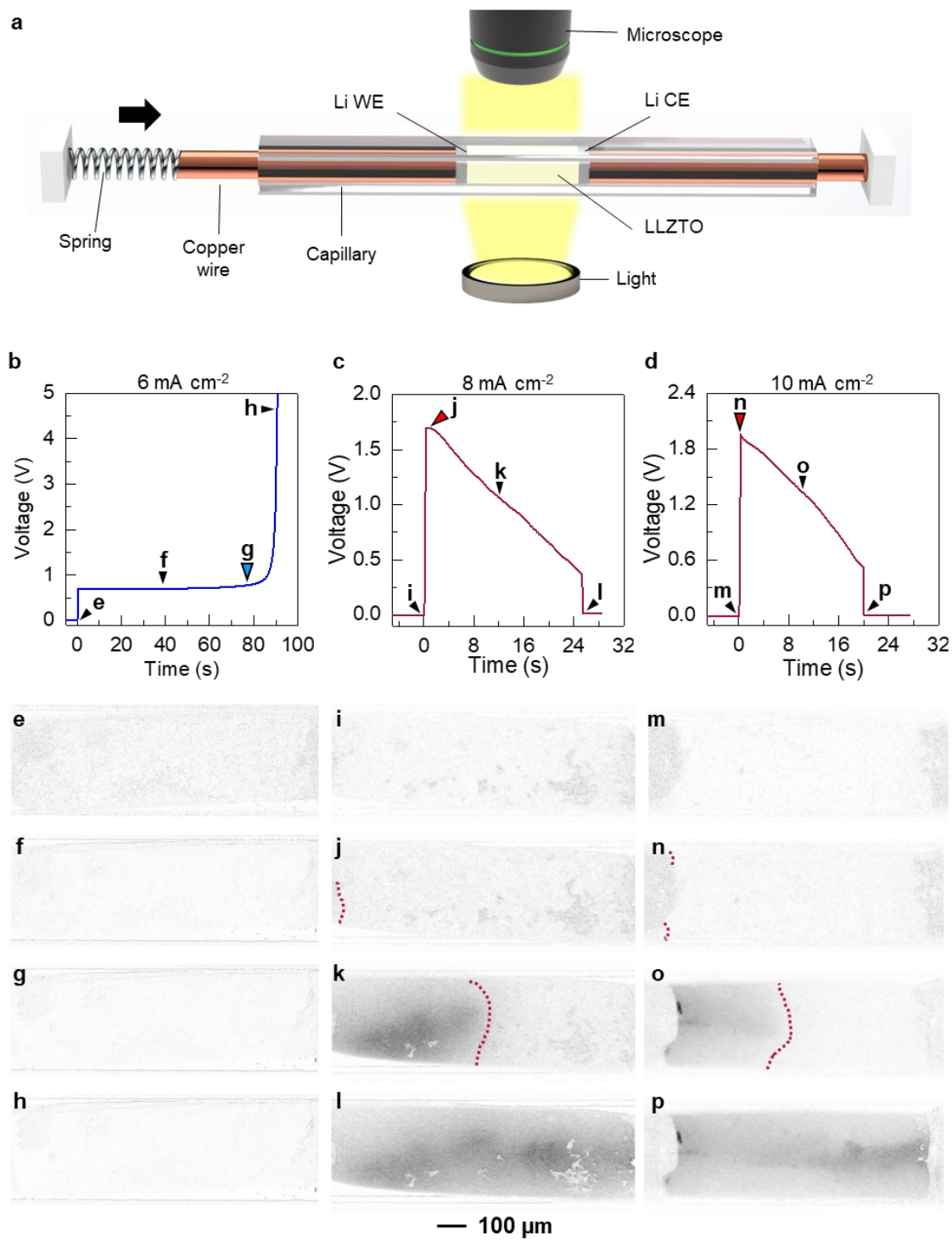

— 100 μm

**Figure 1.** *Operando* **cells determining the influence of current density on LLZTO a,** *Operando* capillary cell setup with a stack pressure of 20 MPa. The working electrode (WE) is on the left with the counter electrode (CE) on the right. A transmission light illuminates the thin opaque SSE piece from below and any visual changes are captured with the microscope lens from above. At different applied current densities, either divergence or penetration is encountered. **b,** At 6 mA cm$^{-2}$ an exponential voltage rise, divergence, occurs. **c,** At 8 mA cm$^{-2}$ a stable voltage plateau followed by a steady decline is encountered, indicating dendritic penetration of the SSE. **d,** At 10 mA cm$^{-2}$ immediate penetration occurs with a lack of any stable voltage plateau. The videos are post-processed into snapshots (See also video S1-S3) **e-p,** Snapshots to highlight any internal growth, the growth front is highlighted by the dashed dark-red lines.

Using increasing current densities, three typical responses were encountered. At a lower current density of 6 mA cm$^{-2}$ (Figure 1b), a stable plateau around 700 mV develops until 50 seconds, at which the voltage gradually increases. After 80 seconds an abrupt increase occurs until the upper voltage cutoff is reached. This gradual to abrupt transition point is marked as the divergence point. The voltage response is referred to as a divergent profile. As shown by the post-processed snapshots (Figure 1e-h) and full-color video comparison (Video S1), no shadows of growth into the electrolyte, whether whisker[27] or dendritic, was observed.

Cells at 8 mA cm$^{-2}$ generated a voltage profile typically encountered for penetrative growth (Figure 1c). While a very short plateau could be identified until 1.1 seconds, the voltage constantly decreases until 25.2 seconds, where it abruptly drops to a negligible value. As seen in Figure 1i, at the start of the experiment, no visible growth occurs, however, as soon as the voltage begins to decrease at 1.2 seconds, a dark region (Video S2 and Figure 1j) evolves at the interface of the working electrode (WE), which moves toward the counter electrode (CE). This dark region is the shadow of dendritic growth[28,29] inside the LLZTO, as the transmitted light is obscured by the metal structures. The initial nucleation occurs at the lower edge of the working electrode, with additional growth expanding through the electrolyte (Video S2). The voltage drop occurs due to a decrease in the inter-electrode distance[30] and an increased deposition area as the dendrite expands through the LLZTO, both of which reduce cell impedance and hence the transient voltage (Figure 1k). Most importantly, this drop in the voltage plateau coincides with the start of visually determined growth, marking this as the significant dendritic initiation point used in later sections. Eventually, the counter electrode (CE) is reached (Figure 1l), culminating in a short circuit and almost negligible voltage. Electrochemical impedance spectroscopy (EIS) further confirms cell penetration due to decreased total impedance (Figure S1). To distinguish from the first case, the voltage response discussed here is referred to as a penetrative profile.

A further increase to 10 mA cm$^{-2}$ leads to a pure penetrative response (Figure 1d), without any stable plateau at the beginning. Instead, an immediate voltage decrease is observed. This is further corroborated by Figure 1n where dendrites initiate instantly upon current application until penetration is complete (Figure 1p and Video S3).

The above three cases appear to suggest the existence of two different growing behaviors of lithium metal. When a low current density is applied, growths of Li metal on the working electrode remain outside of the LLZTO, as confirmed by both the stable voltage plateau and the lack of shadows in LLZTO. The penetration of LLZTO by lithium metal requires the growths to invade the LLZTO, at high enough current densities, the metal structure inside the LLZTO appears to grow at the tip resembling classical dendritic growths of copper[31] and zinc[32] in liquid electrolytes. The existence of the short voltage plateau in the case presented in Figure 1c, i.e. before the voltage decreases and the emergence of shadows inside LLZTO, suggests an "incubation" period for the lithium growth outside of LLZTO to transition into the inward tip growths. Is this incubation time current-dependent? What are the possible connections with the current-dependent Sand's time that marks the transition of root-growing whiskers to tip-growing dendrites in liquid electrolyte?

To explore these 2 behaviors of outward vs inward growth, lithium deposition on a thin copper layer was viewed using an optical microscope. Any deposition of lithium will result in bumps forming on the copper and ultimately piercing it, leading to multiple protrusions on the surface[33,34]. A modified cell was constructed with a 60 nm copper layer sputtered on one surface of a 2×2×2 mm LLZTO sample to act as the working electrode and lithium as the counter. The copper surface was viewed from above (Figure 2a).

As soon as the current is applied, dark spots begin to emerge out of the surface (Figure 2c and Video S4). Some protrusions take on a worm-like appearance with others having a larger flatter coverage (Video S4). Visually the surface growth stops at almost 65 seconds and as shown by the dashed line in Figure 2b, the lithium coverage percentage simultaneously stalls. As in figure 1c, this is an incubation time, indicating any actual growth after is constrained to the electrolyte interior and is the dendritic initiation point (Figure 2d). Even during the progression of internal dendritic growth, the surface changes remain minimal (Figure 2e) and reinforces the exclusive occurrence of an internal growth process at the dendritic initiation point.

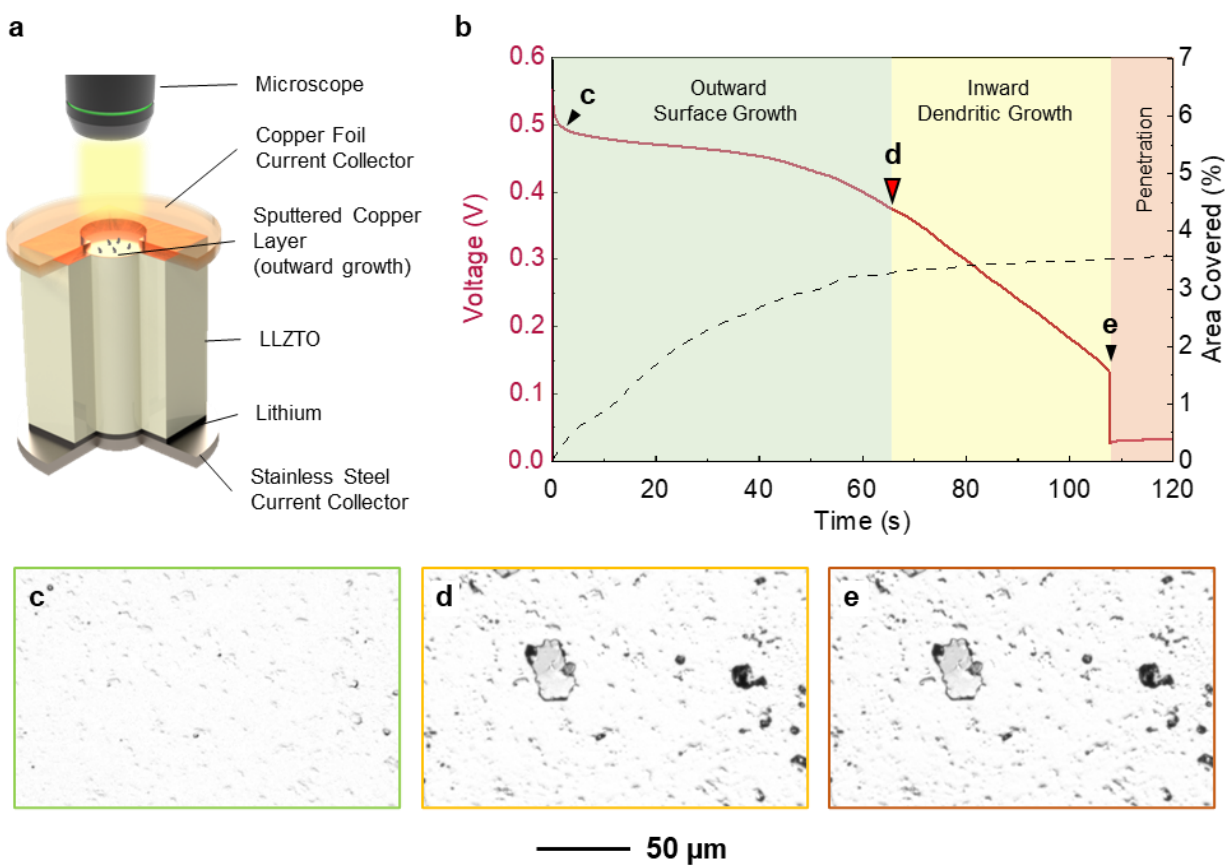

**Figure 2.** *Operando* cell focusing on surface growth out of LLZTO. **a,** Experimental setup of surface growth cell. **b,** Voltage profile at 2 mA cm$^{-2}$ with the modulus of the voltage on the y-axis (See also video S4) **c,** Upon current application small protrusions grow out from the surface. **d,** Protrusions reach a maximum size and stop growing until **e,** internal penetration occurs

## 2.2 Constant Current on Larger Cells

Intrigued by the different profiles shown by the capillary experiments and to enable a higher throughput of cells, systematic experiments were performed with bigger LLZTO pellets. When possible, tests were done on daughter pieces cut from the same mother pellet (Supplementary Figure S2) to ensure high sample consistency. As demonstrated in recent studies[35], a larger cell geometry is more susceptible to pressure variations across the interface due to non-parallelized surfaces or experimental setup error, encouraging void formation in lower pressured regions. Therefore, flat parallel electrode surfaces were ensured to always develop a uniform pressure distribution. Here symmetrical cells with 1-mm-thick LLZTO pellets were initially tested, at current densities ranging from 0.25 mA cm$^{-2}$ to 10 mA cm$^{-2}$.

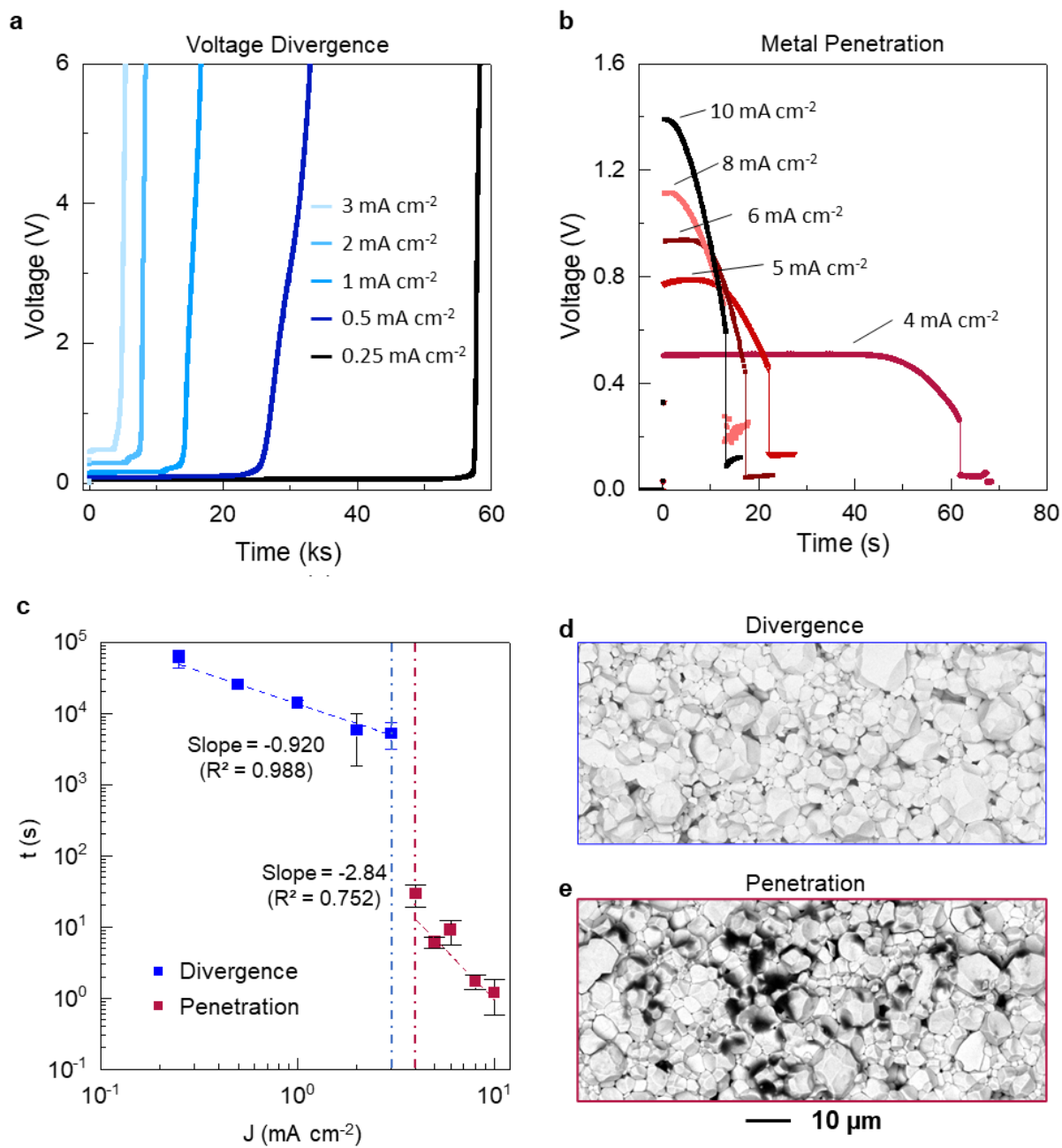

**Figure 3. Constant current on larger cell geometry. a,** For lower current densities divergence is typically encountered. **b,** Higher current densities exhibit penetrative profiles. **c,** Log-log plot of the dendritic initiation and divergence times against current density. A discontinuity exists where either penetration or divergence can occur at the same current density, bounded by the vertical dashed lines. The blue dashed line is the boundary for divergence, whereas the red dashed line is for the penetrative boundary. The error bars show the standard deviation of the data. **d,** SEM sample cross-section for divergent profiles. **e,** SEM sample cross-section for penetrative profiles. Dark bulbous spots show lithium deposits, indicative of internal dendrite growth.

For lower current densities from 0.25 mA cm$^{-2}$ to 3 mA cm$^{-2}$, a divergent voltage profile was typically encountered (Figure 3a). The voltage divergence times versus current densities yield a slope very close to -1 (Figure 3c), suggesting a current-independent accessible capacity. For this divergent region, an average of 4.48 mAh cm$^{-2}$ of lithium is stripped from the interface (Table S1) after which the current can no longer be sustained. This coincides with studies where an accessible capacity[13] was demonstrated, which was affected by electrode thickness and the stripping current density. At very low thicknesses lithium can no longer be effectively driven to the interface[15] while at very low current densities depletion[13] of the CE occurs. It is important to note this accessible capacity, which is higher than capacities determined from the voltage peaks in previous LSV experiments[36] due to the transport limitation within the electrolyte, is tied to the areal capacity of the lithium electrode i.e. a thicker electrode would give a higher accessible capacity.

For higher current densities from 4 mA cm$^{-2}$ to 10 mA cm$^{-2}$, penetrative profiles were mostly encountered (Figure 3b). Seen in Figure 3e, the dendrites appear as dark areas of deposition in the electrolyte[37], with most deposits lying at the intersection of grains, suggesting a growth pathway via the grain boundaries[38]. An upper limit of 10 mA cm$^{-2}$ was chosen to still observe a stable voltage plateau before penetration, as any higher current density would result in immediate penetration similar to Figure 1d with an instantaneous dendritic initiation point. An average of 0.03 mAh cm$^{-2}$ of lithium is stripped, far below the value needed for widespread contact loss at the CE as in the divergent profiles. Interestingly, the log-log plot results in a scaling of -2.84 (Fig 3c), this is noteworthy since a comparable scaling of -2 occurs for transport-limited systems[8,39–41] where at a characteristic time, i.e. Sand's time[42], tip-growing dendrites initiate. This close correlation to -2 and this emergent incubation time resembling a Sand's time suggests the dendritic initiation process in LLZTO is more complex than previously thought and may result due to transport limitations.

The transition from the divergent to the penetrative profiles evolves a discontinuity in the data which starts at the lowest current density for penetration at 4 mA cm$^{-2}$ (red vertical dashed line in Fig. 3c) and ends at the highest current density for divergence at 3 mA cm$^{-2}$ (blue vertical dashed line). The average of these values yields an important new parameter i.e., a *break-through* current density, defined as the lowest current density at which penetration will most likely occur, occurring at 3.5 mA cm$^{-2}$ for 1mm thick samples. This new term differentiates the current from the commonly used critical current density (CCD), which is traditionally found by galvanostatic cycling and not one-way chronopotentiometry as here. To note though, even at the higher 4 mA cm$^{-2}$ density, we did find a single sample showing a divergent behavior with a lack of penetration. This is unsurprising since variation in the SSE[43] and heterogeneity at the surface[44] will lead to no single current value defining when dendrites could initiate, but will rather result in a range of initiation densities which are present in this discontinuity.

### 2.3 Thickness Influence on Penetrative Current Densities

To delve deeper into the underlying mechanisms governing dendrite initiation, variations in the thickness of the solid-state electrolyte (SSE) were investigated. Miniature samples 2 mm and 0.5 mm thick were obtained from a single pellet and tested in a similar manner. To focus primarily on exploring the penetrative region, a current density of 10 mA cm$^{-2}$ was first chosen and then lowered until a divergent profile was encountered.

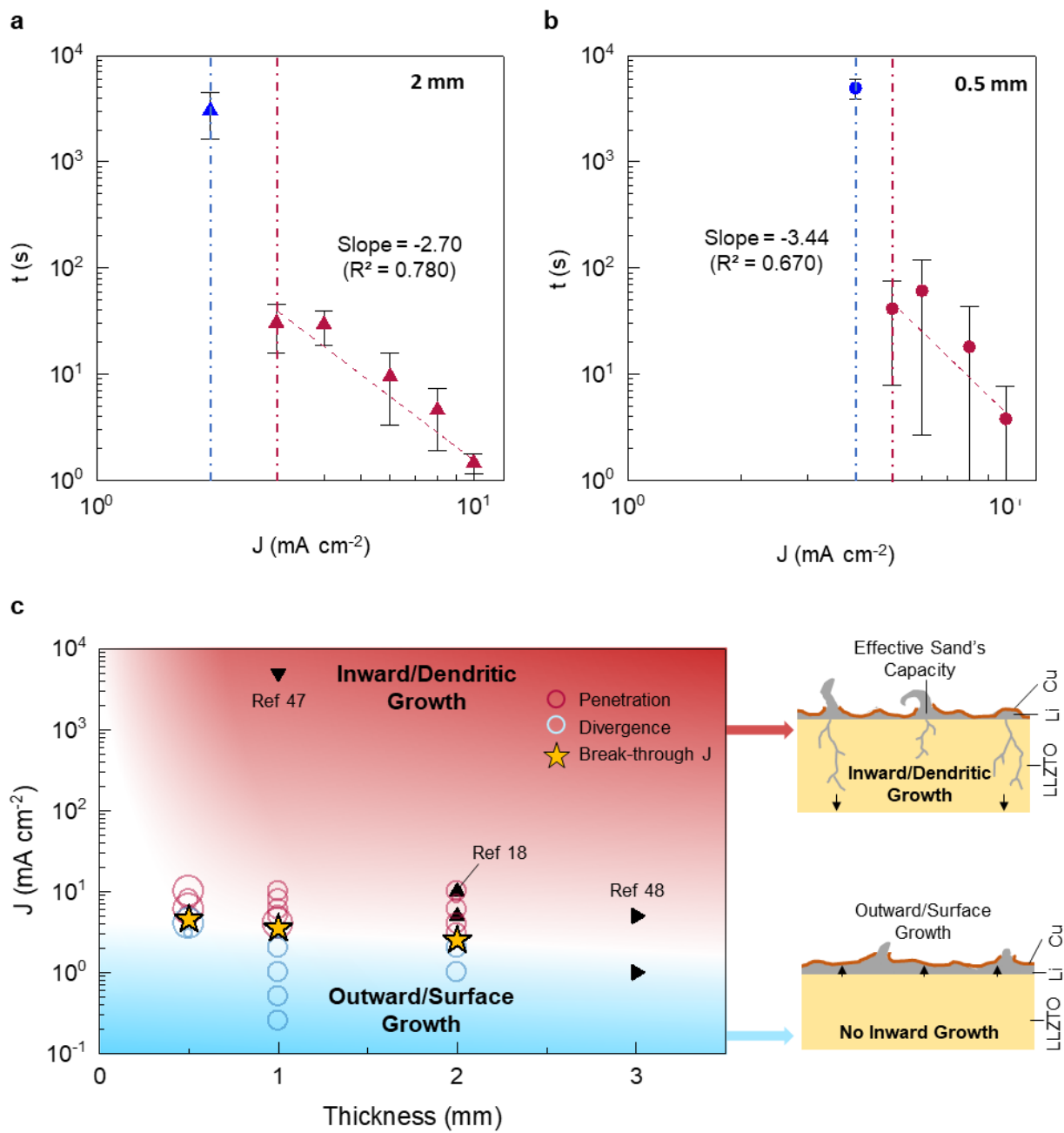

**Figure 4. Log-log plots with standard deviations for samples of varied thicknesses. a,** Thicker 2 mm samples with a -2.3 scaling. **b,** 0.5 mm samples show a slope closer to -3 with larger deviations. **c,** Break-through current vs SSE thickness. The red region shows the dendritic growth zone, whereas the blue region shows a higher likelihood of only surface growth. The varying circle size for the data points corresponds to the number of samples tested. The reference data points shown are all penetrative.

The thicker 2 mm samples shown in Figure 4a exhibited a scaling factor of -2.70, which is closer to the -2 observed in transport-limited systems. Notably, the discontinuity occurred between 2 and 3 mA cm$^{-2}$, with an estimated average of 2.5 mA cm$^{-2}$, indicating a reduced break-through current density with an increase in thickness. The 0.5 mm samples exhibited a scaling of -3.44 (as shown in Figure 4b), similar to the 1 mm samples. The discontinuity occurred at higher current densities, between 4 and 5 mA cm$^{-2}$, with an estimated average of 4.5 mA cm$^{-2}$, indicating an increase in the break-through current density with a reduction in thickness. The inversed relationship between the break-through current density and the thickness we discovered here (Fig. 4c) resembles the well-known thickness dependence of limiting current in liquid electrolytes.

Plotting our own and published data[16–18], the break-through current density we identified set the boundary between two dynamic regimes of lithium growth mechanisms (Fig. 4c). When a high current density is applied, plating of lithium metal within the SSE becomes possible after a short incubation time, yielding a power-law scaling similar to the classic Sand's time. In this case, the penetrative growth of lithium metal is strongly dominated by the chemo-mechanical dynamics of crack formation and propagation. Many groups have investigated this scenario from various perspectives. However, when a low current density is applied, i.e., lower than the break-through current density, lithium growths will simply emerge out of the surface of the SSE and no inward growths are likely to occur.

## 3. Discussion

Despite the many fundamental differences in chemistry, physics, structure, mechanics, etc., between the SSEs and the liquid electrolytes, our *operando* experiments with over 60 highly consistent samples reveal the critical electrochemical limits of SSE that resemble classic behaviors of liquid electrolytes[8,45]. Not only that a high enough current density is required to initiate the inward growth, i.e., the penetrative dendritic lithium growth, but there also exists an "incubation time" that follows a power-law scaling similar to the classic Sand's time scaling for liquid electrolytes. The incubation time can also be seen in a recent experiment by Huang et al.[46], where the nucleation of lithium metal in the LLZO and the propagation of a bowl-shaped crack occurs only after significant and stalled outward protrusion of lithium whiskers out of the tungsten-tip touched platinum pad on the LLZO surface. The bowl-shaped crack coincides with the concentration front[47] of long-range spherical diffusion a microelectrode may induce[41].

Under equilibrium, the ceramic crystal structure with fixed anions and the local electroneutrality constraint preclude significant ion polarization. However, when Faradaic reactions take place on both sides of the LLZO, lithium ions are being continuously shoved in at the positive interface and stripped out at the negative interface, causing lithium-ion enrichment and depletion, respectively. Such departures of stoichiometry are possible due to the intrinsic low occupancy[48] and disordered arrangement[49] of lithium ions on the available lithium sites in cubic LLZO. When a current density higher than the system-specific thickness-dependent break-through current density is applied, long-range transport through SSE becomes insufficient, generating a dynamic space charge layer[50] (SCL) at the negative interface between the SSE and the working electrode[51,52]. Due to the high overpotential to maintain the imposed constant current, the very high electric field strength across this thin SCL could lead to local dielectric breakdown and the direct injection of electrons[53], which will reduce lithium ions that are no longer available at the interface, generating the remote lithium metal clusters[54] that may open up cracks[55] for further penetration of lithium metal toward the counter electrode. However, as we analyzed above, such phenomena do not exist when low current densities are applied.

The use of miniature cells enabled ideally consistent samples, high-throughput characterizations, and more accurate calculation of the true local current density, as the area of the cross-section of the SSE sample is closer to the lithium penetration spot size (Figure S3). However, the use of larger pellets in the top-view operando experiments in Fig. 2 revealed more severe heterogeneity at the interface, yielding local current densities much higher than the averaged current density. Analysis of the interfaces after penetration shows dark spots on both cell surfaces (Figure S3). The true localized current densities were estimated to be 130 mA cm$^{-2}$ and 43 mA cm$^{-2}$ for the large pellet and miniature pellet, respectively. Due to the much higher local current density, lithium dendrite penetrates larger pellets faster than the miniature pellets at the same nominal current density. This is also shown in Figure 2b-e, where lithium protrusions emerged out of a few naturally selected regions and not uniformly across the surface. Without the realization of uniform current distribution, interpretations of the mechanisms of dendrite penetration and critical current densities in SSEs from different studies will remain difficult to reconcile and unify.

In all our cases, the scaling exponent between the "incubation time" and the current density deviated negatively from the classic value of -2 in Sand's formula. Such deviations have been rigorously investigated[45] and are attributed to the geometry of the ion conduction and metal penetration pathway. An expanding pathway will lead to a more negative exponent, e.g., -3.44 in Fig. 4b. Note that the apparent geometry of our LLZTO samples is more or less the same with flat orthogonal faces, it is the collective geometry formed by the naturally selected ion conduction pathways within the electrolyte that determine the dynamics and the incubation/Sand's time. As can be seen in Supplementary Video 2, the penetrative (inward) dendritic growth, as reflected by the shadows, initially occurred at the lower edge of the working electrode (WE) but gradually expanded through the electrolyte. The growth velocities of lithium dendrites in different SSE samples at the same chosen current density are very similar (Figure S4), indicating a similar amount of lithium metal deposits were plated within the samples, confirming highly consistent bulk properties of our samples.

### 4. Conclusions

In this study, we carried out systematic *operando* characterizations of lithium dendrite initiation and growth in more than 60 highly consistent samples of Ta-doped LLZO ceramic electrolytes, using our unique miniature capillary cells for side view observation of inward dendrite growths and coin cells for top-view observation of outward surface growths. Similar to the transition from root-growing whiskers to tip-growing dendrites at over-limiting current densities in liquid electrolytes[8,56], the outward surface growth was found to transition to the inward dendritic growth when the applied current density is higher than the breakthrough current density we identified here, yet only after an incubation time which scales with the applied current density in a trend similar to the Sand's time. The discovery suggests that a transport-limited process, which may be attributed to the formation of the local space charge layer between the SSE and the metal electrode plays a critical role in the initiation of inward dendrite growth. For current densities lower than the breakthrough current density, no inward growths were found, therefore the challenge is mainly to maintain the intimate SSE|Li interfaces. While the thinner SSE tends to allow higher breakthrough current density, which is beneficial for fast charging without inward dendritic growth, maintaining a stable interface becomes increasingly challenging[57,58]. This is evidenced by the large variance observed in our 0.5 mm-thick samples (Fig. 4b). To achieve a desired thickness of 20-50 μm for an acceptable total conductance, ensuring a homogeneous current distribution to avoid high local current density is the key to the success of solid-state batteries.

## 5. Experimental Procedures

*Materials:* Cubic Ta-doped LLZO (LLZTO) with the required composition of $Li_{6.4}La_3Zr_{1.4}Ta_{0.6}O_{12}$ was purchased from MSE Supplies and stored in an inert argon atmosphere to prevent reaction with humid air.

*SSE Fabrication:* To fabricate large numbers of pellets up to 96 rd%, conventional tube furnace sintering was used. The LLZTO powder was mixed with a 2% polyvinyl butyral binder purchased from Fischer Scientific in ethanol. The dried powder was uniaxially pressed at 300 MPa in a 12.7 mm diameter stainless-steel die. This green body was then isostatically pressed at 350 MPa for 10 mins. For sintering, the green body was suspended on a platinum support[23] and placed in a magnesia (MgO) crucible covered with an accompanying lid. 0.5 g of loose LLZTO powder was placed in the crucible as bed powder and sintered in a tube furnace at 1250 °C for 20 mins. A ramp rate of 3 °C min$^{-1}$ was used for the heating and cooling process. The final pellets had a tan coloration.

For larger samples referred to in the text, the sintered pellets were then cut using a low-speed diamond saw into multiple pieces. These pieces were polished using 1200 grit and 2500 grit sandpaper followed by fine polishing on a polishing pad with a 50 nm alumina glycol solution. The samples were rinsed with isopropyl alcohol and immediately placed in an argon-filled glove box to minimize the surface reaction with air[59].

For the capillary cells, these cut pieces were systematically ground down using sandpaper until the desired geometry could fit into a 400 μm width square capillary. These samples were rinsed with isopropyl alcohol and immediately placed in an argon environment.

*Lithium Electrode:* The cut samples were rubbed over molten lithium (175 °C) until the entire interface was covered. For the capillary cells, the same procedure was employed, except specially made tweezers were made to allow for handling the very tiny samples.

*Sputtered copper electrode:* The SSE was masked with Kapton tape and sputtered using the Kurt Lesker PVD 75 Physical Vapor Deposition system at a rate of 0.5 nm s$^{-1}$.

*SEM Imaging:* Thermofisher Quattro S environmental SEM was used at 10 kV accelerating voltage and 10 mm working distance. The polished top surfaces of the LLZTO and Li|SSE interface were transferred to the SEM in an airtight homemade SEM holder to minimize exposure to air and subsequent oxidation.

*Optical imaging:* An Olympus BX53M microscope was used for all *operando* tests. The captured was post-processed using ImageJ, where a macro was written to enable frame-by-frame comparison. Any changes in the pixels resulted in a grey color output, which could be used to extract dendritic growth, referred to as the pixel intensity difference.

*Electrochemical Measurements*: A Gamry Reference 600+ was used for the EIS and constant current measurements. An amplitude of 10mV with 8 points/decade was used for the data collection, with a frequency range of 5 Mhz to 0.005.

**Supplemental Information**

Supplemental Information can be found online at EES.

**Supplementary Videos**

**Video S1** – operando capillary cell at 6 mA cm$^{-2}$

**Video S2** – operando capillary cell at 8 mA cm$^{-2}$

**Video S3** – operando capillary cell at 10 mA cm$^{-2}$

**Video S4** – operando surface growth cell at 2 mA cm$^{-2}$

**Data availability**

The data that support the findings of this study are available from the corresponding author upon reasonable request.

**Acknowledgments**

This work is supported by a National Science Foundation Grant (Award No. 2203994). The materials characterization experiments were partially supported by IMSE (Institute of Materials Science and Engineering) at Washington University in St. Louis. The authors thank Dr. James Schilling and Dr. James Buckley for their help with the diamond saw.

**Author contributions**

P.B. designed and supervised the study and led the theoretical analysis. R.G. sintered the LLZTO, fabricated the cells, performed electrochemical tests and microscopy characterization, and analyzed the data. P.B. and R.G. wrote and revised the manuscript.

**Declaration of Interests**

The authors declare no competing financial interests.

## 6. References

1    M. Jia, N. Zhao, H. Huo and X. Guo, *Comprehensive Investigation into Garnet Electrolytes Toward Application-Oriented Solid Lithium Batteries*, Springer Singapore, 2020, vol. 3.

2    S. Yubuchi, M. Uematsu, C. Hotehama, A. Sakuda, A. Hayashi and M. Tatsumisago, *J. Mater. Chem. A*, 2019, **7**, 558–566.

3    Y. Kato, S. Hori, T. Saito, K. Suzuki, M. Hirayama, A. Mitsui, M. Yonemura, H. Iba and R. Kanno, *Nat. Energy*, 2016, **1**, 1–7.

4    T. Asano, A. Sakai, S. Ouchi, M. Sakaida, A. Miyazaki and S. Hasegawa, *Adv. Mater.*, , DOI:10.1002/adma.201803075.

5    F. Zheng, M. Kotobuki, S. Song, M. O. Lai and L. Lu, *J. Power Sources*, , DOI:10.1016/j.jpowsour.2018.04.022.

6    V. Thangadurai, D. Pinzaru, S. Narayanan and A. K. Baral, *J. Phys. Chem. Lett.*, 2015, **6**, 292–299.

7    A. J. Samson, K. Hofstetter, S. Bag and V. Thangadurai, *Energy Environ. Sci.*, 2019, **12**, 2957–2975.

8    P. Bai, J. Li, F. R. Brushett and M. Z. Bazant, *Energy Environ. Sci.*, 2016, **9**, 3221–3229.

9    A. Sharafi, C. G. Haslam, R. D. Kerns, J. Wolfenstine and J. Sakamoto, *J. Mater. Chem. A*, 2017, **5**, 21491–21504.

10    Q. Zhao, S. Stalin, C. Z. Zhao and L. A. Archer, *Nat. Rev. Mater.*, 2020, **5**, 229–252.

11    Y. Lu, C. Z. Zhao, H. Yuan, X. B. Cheng, J. Q. Huang and Q. Zhang, *Adv. Funct. Mater.*, 2021, **2009925**, 1–33.

12    J. Kasemchainan, S. Zekoll, D. Spencer Jolly, Z. Ning, G. O. Hartley, J. Marrow and P. G. Bruce,

*Nat. Mater.*, 2019, **18**, 1105–1111.

13  K. Lee, E. Kazyak, M. J. Wang, N. P. Dasgupta and J. Sakamoto, *Matter*, , DOI:10.1016/j.joule.2022.09.009.

14  M. J. Wang, R. Choudhury and J. Sakamoto, *Joule*, 2019, **3**, 2165–2178.

15  C. G. Haslam, J. B. Wolfenstine and J. Sakamoto, *J. Power Sources*, 2022, **520**, 230831.

16  T. Swamy, R. Park, B. W. Sheldon, D. Rettenwander, L. Porz, S. Berendts, R. Uecker, W. C. Carter and Y.-M. Chiang, *J. Electrochem. Soc.*, 2018, **165**, A3648–A3655.

17  L. Porz, T. Swamy, B. W. Sheldon, D. Rettenwander, T. Frömling, H. L. Thaman, S. Berendts, R. Uecker, W. C. Carter and Y. M. Chiang, *Adv. Energy Mater.*, , DOI:10.1002/aenm.201701003.

18  G. McConohy, X. Xu, T. Cui, E. Barks, S. Wang, E. Kaeli, C. Melamed, X. W. Gu and W. C. Chueh, *Nat. Energy*, 2023, **8**, 241–250.

19  X. Liu, R. Garcia-Mendez, A. R. Lupini, Y. Cheng, Z. D. Hood, F. Han, A. Sharafi, J. C. Idrobo, N. J. Dudney, C. Wang, C. Ma, J. Sakamoto and M. Chi, *Nat. Mater.*, , DOI:10.1038/s41563-021-01019-x.

20  F. Han, A. S. Westover, J. Yue, X. Fan, F. Wang, M. Chi, D. N. Leonard, N. J. Dudney, H. Wang and C. Wang, *Nat. Energy*, 2019, **4**, 187–196.

21  K. K. Fu, Y. Gong, G. T. Hitz, D. W. Mcowen, Y. Li, S. Xu, Y. Wen, L. Zhang, C. Wang and G. Pastel, 2017, **12**, 1568–1575.

22  G. T. Hitz, D. W. McOwen, L. Zhang, Z. Ma, Z. Fu, Y. Wen, Y. Gong, J. Dai, T. R. Hamann, L. Hu and E. D. Wachsman, *Mater. Today*, 2019, **22**, 50–57.

23  X. Huang, Y. Lu, H. Guo, Z. Song, T. Xiu, M. E. Badding and Z. Wen, *ACS Appl. Energy Mater.*, 2018, **1**, 5355–5365.

24   T. Krauskopf, B. Mogwitz, C. Rosenbach, W. G. Zeier and J. Janek, *Adv. Energy Mater.*, , DOI:10.1002/aenm.201902568.

25   T. Krauskopf, H. Hartmann, W. G. Zeier and J. Janek, *ACS Appl. Mater. Interfaces*, 2019, **11**, 14463–14477.

26   X. Zhang, Q. J. Wang, K. L. Harrison, S. A. Roberts and S. J. Harris, *Cell Reports Phys. Sci.*, 2020, **1**, 100012.

27   T. Krauskopf, R. Dippel, H. Hartmann, K. Peppler, B. Mogwitz, F. H. Richter, W. G. Zeier and J. Janek, *Joule*, 2019, **3**, 2030–2049.

28   E. Kazyak, S. William, C. Haslam, J. Sakamoto, N. P. Dasgupta, E. Kazyak, R. Garcia-mendez, W. S. Lepage, A. Sharafi, A. L. Davis, A. J. Sanchez, K. Chen, C. Haslam, J. Sakamoto and N. P. Dasgupta, *Matter*, 2020, **2**, 1025–1048.

29   W. Guo, F. Shen, J. Liu, Q. Zhang, H. Guo, Y. Yin, J. Gao, Z. Sun, X. Han and Y. Hu, *Energy Storage Mater.*, 2021, **41**, 791–797.

30   L. E. Marbella, S. Zekoll, J. Kasemchainan, S. P. Emge, P. G. Bruce and C. P. Grey, *Chem. Mater.*, 2019, **31**, 2762–2769.

31   J. H. Han, E. Khoo, P. Bai and M. Z. Bazant, *Sci. Rep.*, 2014, **4**, 1–8.

32   Y. Lee and P. Bai, *J. Electrochem. Soc.*, 2023, **170**, 060511.

33   M. Motoyama, M. Ejiri and Y. Iriyama, *J. Electrochem. Soc.*, 2015, **162**, A7067–A7071.

34   M. Motoyama, Y. Tanaka, T. Yamamoto, N. Tsuchimine, S. Kobayashi and Y. Iriyama, *ACS Appl. Energy Mater.*, 2019, **2**, 6720–6731.

35   E. Kazyak, M. J. Wang, K. Lee, S. Yadavalli, A. J. Sanchez, M. D. Thouless, J. Sakamoto and N. P. Dasgupta, *Matter*, 2022, 1–23.

36   R. Gopal, L. Wu, Y. Lee, J. Guo and P. Bai, *ACS Energy Lett.*, 2023, **8**, 2141–2149.

37   Y. Ren, Y. Shen, Y. Lin and C. W. Nan, *Electrochem. commun.*, 2015, **57**, 27–30.

38   H. Huo, J. Gao, N. Zhao, D. Zhang, N. G. Holmes, X. Li, Y. Sun, J. Fu, R. Li, X. Guo and X. Sun, *Nat. Commun.*, 2021, **12**, 1–10.

39   T. F. Fuller and J. N. Harb, *Electrochemical Engineering.*, 2018.

40   J. Newman and K. E. Thomas-alyea, *wiley*.

41   A. Bard and L. Faulkner, *Electrochemical methods: fundamentals and applications*, 2001, vol. 2.

42   H. J. S. Sand, *London, Edinburgh, Dublin Philos. Mag. J. Sci.*, 1901, **1**, 45–79.

43   A. Wachter-Welzl, J. Kirowitz, R. Wagner, S. Smetaczek, G. C. Brunauer, M. Bonta, D. Rettenwander, S. Taibl, A. Limbeck, G. Amthauer and J. Fleig, *Solid State Ionics*, 2018, **319**, 203–208.

44   A. Wachter-Welzl, R. Wagner, D. Rettenwander, S. Taibl, G. Amthauer and J. Fleig, *J. Electroceramics*, 2017, **38**, 176–181.

45   Y. Lee, B. Ma and P. Bai, *Energy Environ. Sci.*, 2020, **13**, 3504–3513.

46   J. Zhao, Y. Tang, Q. Dai, C. Du, Y. Zhang, D. Xue, T. Chen, J. Chen, B. Wang, J. Yao, N. Zhao, Y. Li, S. Xia, X. Guo, S. J. Harris, L. Zhang, S. Zhang, T. Zhu and J. Huang, 2022, 524–532.

47   Y. Zhang, Y. Dong and J. Li, , DOI:10.1016/j.actamat.2023.119620.

48   K. Meier, T. Laino and A. Curioni, .

49   J. Holland, T. Demeyere, A. Bhandari, F. Hanke and V. Milman, 2023, 8–13.

50   N. J. J. De Klerk and M. Wagemaker, *ACS Appl. Energy Mater.*, 2018, **1**, 5609–5618.

51   K. Yamamoto, Y. Iriyama and T. Hirayama, *Microscopy*, 2017, **66**, 50–61.

52  K. Yamamoto, Y. Iriyama, T. Asaka, T. Hirayama, H. Fujita, C. A. J. Fisher, K. Nonaka, Y. Sugita and Z. Ogumi, *Angew. Chemie - Int. Ed.*, 2010, **49**, 4414–4417.

53  Y. Aizawa, K. Yamamoto, T. Sato, H. Murata, R. Yoshida, C. A. J. Fisher, T. Kato, Y. Iriyama and T. Hirayama, *Ultramicroscopy*, 2017, **178**, 20–26.

54  X. Liu, R. Garcia-Mendez, A. R. Lupini, Y. Cheng, Z. D. Hood, F. Han, A. Sharafi, J. C. Idrobo, N. J. Dudney, C. Wang, C. Ma, J. Sakamoto and M. Chi, *Nat. Mater.*, , DOI:10.1038/s41563-021-01019-x.

55  Z. Ning, G. Li, D. L. R. Melvin, Y. Chen, J. Bu, D. Spencer-jolly, J. Liu, B. Hu, X. Gao, J. Perera, C. Gong, S. D. Pu, S. Zhang, B. Liu, G. O. Hartley, A. J. Bodey, R. I. Todd, P. S. Grant, D. E. J. Armstrong, T. J. Marrow, C. W. Monroe and P. G. Bruce, , DOI:10.1038/s41586-023-05970-4.

56  B. Ma and P. Bai, *Adv. Energy Mater.*, 2022, **12**, 1–9.

57  H. Hao, Y. Liu, S. M. Greene, G. Yang, K. G. Naik, B. S. Vishnugopi, Y. Wang, H. Celio, A. Dolocan, W. Y. Tsai, R. Fang, J. Watt, P. P. Mukherjee, D. J. Siegel and D. Mitlin, *Adv. Energy Mater.*, 2023, **2301338**, 1–16.

58  Y. Wang, Y. Liu, M. Nguyen, J. Cho, N. Katyal, B. S. Vishnugopi, H. Hao, R. Fang, N. Wu, P. Liu, P. P. Mukherjee, J. Nanda, G. Henkelman, J. Watt and D. Mitlin, *Adv. Mater.*, 2023, **35**, 1–17.

59  A. Sharafi, S. Yu, M. Naguib, M. Lee, C. Ma, H. M. Meyer, J. Nanda, M. Chi, D. J. Siegel and J. Sakamoto, *J. Mater. Chem. A*, 2017, **5**, 13475–13487.

# Supplementary Information

# Intrinsic Electrochemical Limits Preceding Dendrite Penetration in Ceramic Electrolytes

*Rajeev Gopal[a] and Peng Bai[a,b,*]*

[a] Department of Energy, Environmental and Chemical Engineering, Washington University in St. Louis, 1 Brooking Dr, St. Louis, MO 63130, USA.

[b] Institute of Materials Science and Engineering, Washington University in St. Louis, 1 Brooking Dr, St. Louis, MO 63130, USA.

* Correspondence to: pbai@wustl.edu

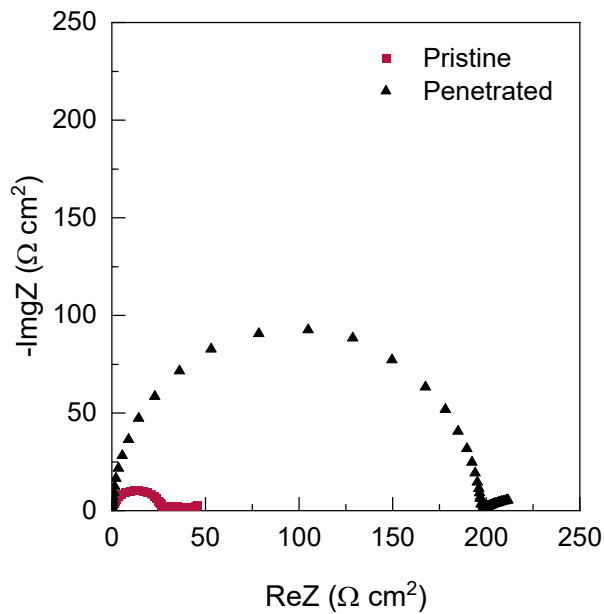

**Figure S1. Nyquist plots of capillary cell before after penetration.** Using EIS a capillary cell is tested before and after a constant current of 10 mA cm$^{-2}$ is applied. The penetrated cell has significantly lower impedance due to the internal growth in the electrolyte, increasing the total WE area and decreasing the inter-electrode distance. While the penetrated sample should lead to a collapse in the capacitive semi-circle and give an agglomeration of points clustering around a low resistance value, a smaller semi-circle is instead developed. Unavoidable extending wires in the experimental setup will invariably lead to the introduction of capacitance in the system. Additionally, despite the dendrite spanning the length of the SSE, soft-shorts could be made, where the tip of the dendrite closest to the WE is destroyed with the sudden influx in current. This effectively creates a very thin SSE which exhibits a typical capacitive impedance response.

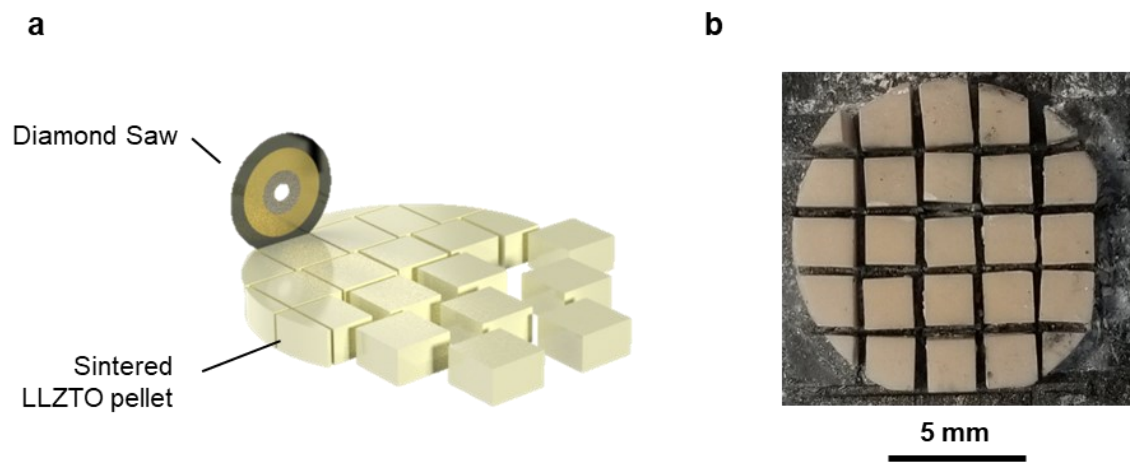

**Figure S2. Preparation of uniform LLZTO samples. a** A singular parent pellet is cut with a diamond saw into smaller children samples[1]. This removes any potential sample differences. **b** Photo of the final cut pellet.

**Table S1. Average areal capacity at penetration or divergence for Fig. 2**

| Current Density (mA cm$^{-2}$) | Areal Capacity (mAh cm$^{-2}$) | Profile |
|---|---|---|
| 0.25 | 4.46 | Divergent |
| 0.5 | 4.19 | |
| 1 | 4.54 | |
| 2 | 3.98 | |
| 3 | 5.85 | |
| 4 | 0.0564 | Penetrative |
| 5 | 0.0302 | |
| 6 | 0.035 | |
| 8 | 0.0277 | |
| 10 | 0.0317 | |

The areal capacity shifts from a relatively large value of over 4 mAh cm$^{-2}$ to below 0.05 mAh cm$^{-2}$ as the voltage profile transitions from a divergent to penetrative response. This indicates the interface, more specifically the counter electrode, is not the limiting factor encouraging dendrite growth in the penetrative region.

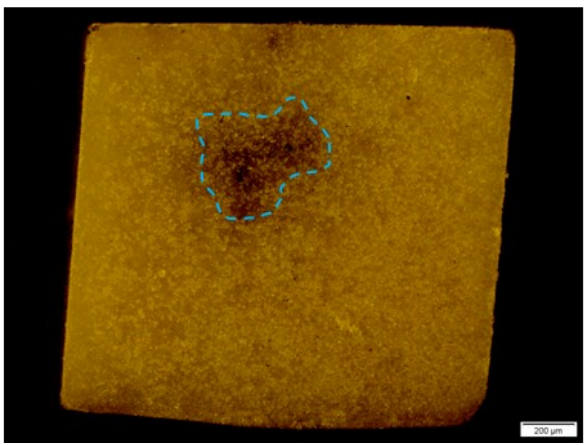 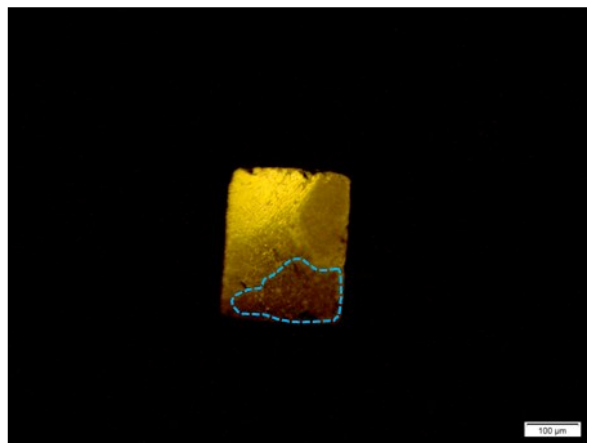

**Figure S3. Optical microscope images of interfaces after penetration at 10 mA cm$^{-2}$.** The blue dashed line shows the area used to estimate a localized current density. **a,** Larger samples show a localized current density of 130 mA cm$^{-2}$ whereas the **b,** capillary samples exhibit 43 mA cm$^{-2}$

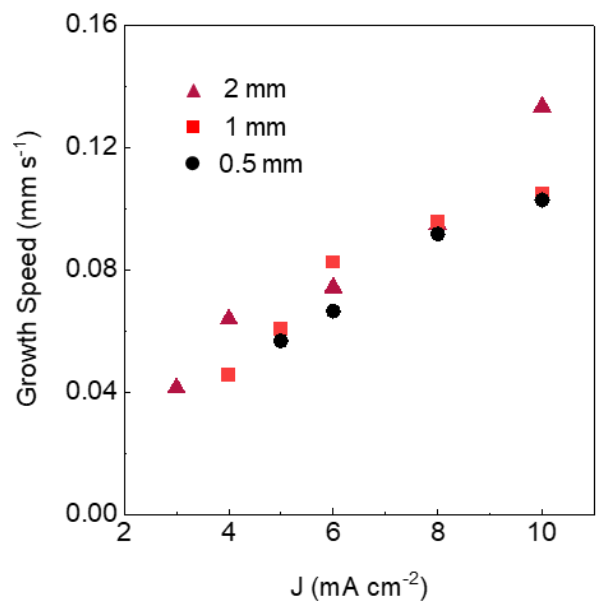

**Figure S4. Average dendrite growth speed at each current density.** Here, while the dendritic growth pattern has expanding tree-like structure such as in the supplementary videos, a linear growth path was assumed to allow an initial comparison of velocities. With the exception of the 2 mm samples at 10 mA cm$^{-2}$, the data show the growth speed at each current density is the similar regardless of thickness, showing the tested samples have very consistent bulk properties.

References

1    R. Gopal, L. Wu, Y. Lee, J. Guo and P. Bai, *ACS Energy Lett.*, 2023, **8**, 2141–2149.